\documentclass[lettersize,journal]{IEEEtran}
\usepackage{amsmath,amsfonts}
\usepackage{algorithmic}
\usepackage{algorithm}
\usepackage{array}
\usepackage[caption=false,font=normalsize,labelfont=sf,textfont=sf]{subfig}
\usepackage{textcomp}
\usepackage{stfloats}
\usepackage{url}
\usepackage{verbatim}
\usepackage{graphicx}
\usepackage{balance}
\usepackage{cite}

\begin{document}

\title{\fontsize{22}{26}\selectfont OTA Characterization of Dual-User IEEE 802.11be EHT-MU Under Transmit-Chain Imbalance%
\thanks{This work was supported by the UKRI/EPSRC Prosperity Partnership in Secure Wireless Agile Networks (SWAN), Grant EP/T005572/1, and the EPSRC Hub in All Spectrum Connectivity: Successive Interference Cancellation for Dynamic Spectrum Access (SINATRA), Grant EP/Y037197/1}
}

\author{%
\IEEEauthorblockN{Mir Lodro, Francesco Raimondo, Geoffrey S. Hilton, Mark A. Beach, and Andrew C. M. Austin\\}
\IEEEauthorblockA{
Communication Systems and Networks Research Group\\
School of Electrical, Electronic and Mechanical Engineering\\
University of Bristol\\
Bristol, United Kingdom\\
mir.lodro@bristol.ac.uk}
}

% ICEAA-IEEE APWC: DO NOT USE \markboth

% ICEAA-IEEE APWC: DO NOT USE \IEEEpubid

% ICEAA-IEEE APWC: DO NOT USE PACE NUMERING 
\pagenumbering{gobble}

\maketitle

\begin{abstract}
This paper presents a controlled over-the-air (OTA) characterization of dual-user IEEE 802.11be Extremely High Throughput Multi-User (EHT-MU) transmission under transmit-chain imbalance. The objective is to determine whether attenuation applied to one access-point transmit chain produces packet-global degradation or appears primarily as stream-dependent payload degradation after receiver processing. Measurements are performed in a shielded RF enclosure using two NI USRP-2953R and NI USRP-2942R software-defined radios, with one USRP generating a dual-user non-OFDMA EHT-MU waveform and the other implementing synchronized dual-branch packet recovery. A calibrated attenuation sweep is applied to the second AP transmit chain (TX2), and performance is evaluated using bit error rate (BER), EHT-Data error vector magnitude (EVM), control-field success probability, payload-success probability, and subcarrier-level EVM distributions. The results show that the stream decoded as User~1 remains at the BER floor over the tested range, while the stream decoded as User~2 exhibits progressive EVM degradation followed by threshold-like BER and payload-success collapse. Common signaling fields remain recoverable, indicating that the dominant observed failure mode is stream-local at the receiver output rather than the packet-global. Replacing User~2 binary convolutional coding (BCC) with low density parity check (LDPC) coding delays the BER and payload-success collapse by approximately \(5\)~dB of TX2 attenuation, demonstrating a measurable coding-dependent robustness margin for the more sensitive stream.
\end{abstract}

\begin{IEEEkeywords}
IEEE 802.11be, Wi-Fi 7, EHT-MU, MU-MIMO, non-OFDMA, OTA measurements, transmit-chain imbalance, BER, EVM, SDR.
\end{IEEEkeywords}

\section{Introduction}
IEEE 802.11be, also known as Wi-Fi~7, extends WLAN capability through Extremely High Throughput (EHT) features including enhanced multi-user operation, larger bandwidth options, and higher spectral efficiency \cite{ieee80211wg,deng2020wifi7,mathworks_wifi7}. Among these features, multi-user transmission is particularly important because it allows an access point (AP) to serve multiple users within a shared packet structure. This improves spectral efficiency, but also makes packet recovery sensitive to spatial imbalance, unequal channel conditions, and hardware-chain asymmetries. In a downlink EHT-MU transmission, common packet fields and user-specific data fields coexist within the same waveform. A practical question is therefore whether an impairment affecting one transmit or spatial dimension causes packet-global failure, or whether the receiver continues to recover the common signaling while degradation appears primarily in one recovered user stream. This distinction is important because packet-global failure would indicate loss of synchronization, format detection, or common signaling, whereas stream-local degradation would indicate that the packet remains interpretable, but one user's payload becomes unreliable. Although the PHY and MAC features of IEEE 802.11be are now well documented \cite{deng2020wifi7,lopez2022mlo,adame2021tsn,murti2021mlo}, comparatively few studies have conducted controlled OTA assessments of dual-user EHT-MU robustness under transmit-chain or stream-dependent imbalance. Such measurements are valuable because they capture practical effects, including synchronization sensitivity, RF front-end nonidealities, spatial-channel conditioning, and packet-recovery behavior that are not always evident in idealized baseband analysis. In this paper, we study a dual-user IEEE 802.11be EHT-MU transmission under progressively increasing attenuation applied to the second AP transmit chain (TX2). The considered waveform is a dual-user \emph{non-OFDMA} EHT-MU transmission with two AP transmit chains and one spatial stream per user. Although the imposed attenuation is transmit-chain-based rather than directly applied to a user receive path, the measured degradation appears primarily in the stream decoded as User~2. This makes the setup suitable for examining stream-dependent degradation, shared-field robustness, and coding-dependent differences in payload recovery. The main contributions of this paper are as follows:
\begin{itemize}
    \item A repeatable shielded-enclosure OTA methodology for dual-user non-OFDMA IEEE 802.11be EHT-MU measurements using NI USRP-2953R and NI USRP-2942R software-defined radios.
    \item A measurement-based characterization of stream-dependent degradation under controlled TX2 transmit-chain attenuation.
    \item A field-wise analysis separating packet-stage signaling recovery from user-specific EHT-Data recovery using BER, EVM, control-field success probability, and payload-success probability.
    \item A coding comparison showing that User~2 LDPC provides an approximately \(5\)~dB TX2-attenuation margin advantage over User~2 BCC in the measured setup.
   
\end{itemize}
The remainder of the paper is structured as follows. Section~II introduces the technical background and the dual-user downlink system model. Section~III describes the experimental setup, waveform generation, receiver processing chain, and performance metrics. Section~IV presents the measurement results and discusses the effects of TX2 attenuation and coding choice. Section~V concludes the paper.

\section{Technical Background}

\subsection{IEEE 802.11be EHT-MU Packet Structure}
IEEE 802.11be retains legacy preamble fields such as the Legacy Short Training Field (L-STF), Legacy Long Training Field (L-LTF), and Legacy Signal field (L-SIG), while adding Extremely High Throughput (EHT)-specific fields including the Universal Signal field (U-SIG), EHT Signal field (EHT-SIG), EHT Long Training Field (EHT-LTF), and EHT Data field (EHT-Data) \cite{deng2020wifi7,mathworks_wifi7}. These fields are useful not only for receiver processing but also for failure localization. If common fields remain decodable while one user stream degrades, the impairment mechanism can be classified as stream-local rather than packet-global. The transmission considered in this paper is a dual-user \emph{non-OFDMA} EHT-MU packet. The measurement campaign does not study separate OFDMA resource-unit allocation across users. Instead, the receiver identifies the packet as a downlink EHT-MU transmission with two spatial streams. The considered configuration is a two-user downlink MU-MIMO system. The AP employs two transmit chains and transmits two spatial streams within the same EHT-MU packet, with one spatial stream assigned to each single-antenna user. Thus, the multi-antenna transmission is used for spatial multiplexing across users, rather than transmit diversity. In the measurement platform, the two receive RF chains are implemented on the same USRP for synchronization and logging convenience. Logically, however, they represent two single-antenna users, and the EHT-MU receiver processes one spatial stream per user.

\subsection{MIMO System Model}
The baseband signal model for the dual-user downlink transmission can be written as
\begin{equation}
\mathbf{y}[n] = \sum_{\ell=0}^{L-1}\mathbf{H}_{\ell}\mathbf{x}[n-\ell] + \mathbf{w}[n],
\end{equation}
where \(\mathbf{x}[n]\in\mathbb{C}^{N_t\times 1}\) is the transmitted sample vector across the \(N_t\) AP transmit chains, \(\mathbf{y}[n]\in\mathbb{C}^{N_r\times 1}\) is the received sample vector across the receive branches, \(\mathbf{H}_{\ell}\) is the \(\ell\)-th MIMO channel tap matrix, \(L\) is the channel memory, and \(\mathbf{w}[n]\) is additive noise. For a given OFDM subcarrier \(k\), the transmitted antenna-domain signal can be written as
\begin{equation}
\mathbf{x}_{\mathrm{tx}}[k,m]=\mathbf{Q}[k]\mathbf{s}[k,m],
\end{equation}
where
\begin{equation}
\mathbf{s}[k,m]=
\begin{bmatrix}
s_1[k,m] \\
s_2[k,m]
\end{bmatrix}
\end{equation}
contains the two user streams, and \(\mathbf{Q}[k]\) denotes the spatial mapping or precoding matrix. In the presence of attenuation on the second AP transmit chain, the effective transmitted signal becomes
\begin{equation}
\mathbf{x}'_{\mathrm{tx}}[k,m]=\mathbf{D}_{\mathrm{TX2}}(A)\mathbf{Q}[k]\mathbf{s}[k,m],
\end{equation}
where
\begin{equation}
\mathbf{D}_{\mathrm{TX2}}(A)=
\begin{bmatrix}
1 & 0 \\
0 & \alpha(A)
\end{bmatrix},
\end{equation}
and \(0\leq \alpha(A)\leq 1\) represents the linear attenuation factor corresponding to the applied TX2 attenuation \(A\). The received OFDM symbol vector is then
\begin{equation}
\mathbf{Y}[k,m]=\mathbf{H}[k]\mathbf{D}_{\mathrm{TX2}}(A)\mathbf{Q}[k]\mathbf{s}[k,m]+\mathbf{W}[k,m].
\end{equation}
Equivalently, the attenuation modifies the effective MIMO channel as
\begin{equation}
\mathbf{H}_{\mathrm{eff}}[k,A]=\mathbf{H}[k]\mathbf{D}_{\mathrm{TX2}}(A)\mathbf{Q}[k].
\end{equation}
Each receive branch therefore observes a superposition of the signals radiated by both AP transmit antennas. The receiver identifies the user allocation from the EHT-MU signaling fields and estimates the effective spatial-stream channel from the EHT-LTF. Spatial equalization then separates the transmitted streams:
\begin{equation}
\hat{\mathbf{s}}[k,m]=\mathbf{G}[k]\mathbf{Y}[k,m],
\end{equation}
where \(\mathbf{G}[k]\) denotes the equalization matrix, for example, zero-forcing or minimum mean square (MMSE) equalization. The recovered streams are then mapped back to User~1 and User~2 according to the known EHT-MU configuration and decoded signaling information. In practice, coarse and fine carrier-frequency-offset correction, timing refinement, and pilot-based tracking are necessary before reliable data recovery can be achieved. These processing stages motivate the field-wise diagnostics used in the measurements. Failures before or during U-SIG and EHT-SIG recovery indicate packet-stage impairment, whereas successful signaling recovery followed by user-specific BER or EVM degradation indicates stream-local payload impairment.
\section{Experimental Setup}
\begin{figure}[t]
    \centering
    \includegraphics[width=\columnwidth]{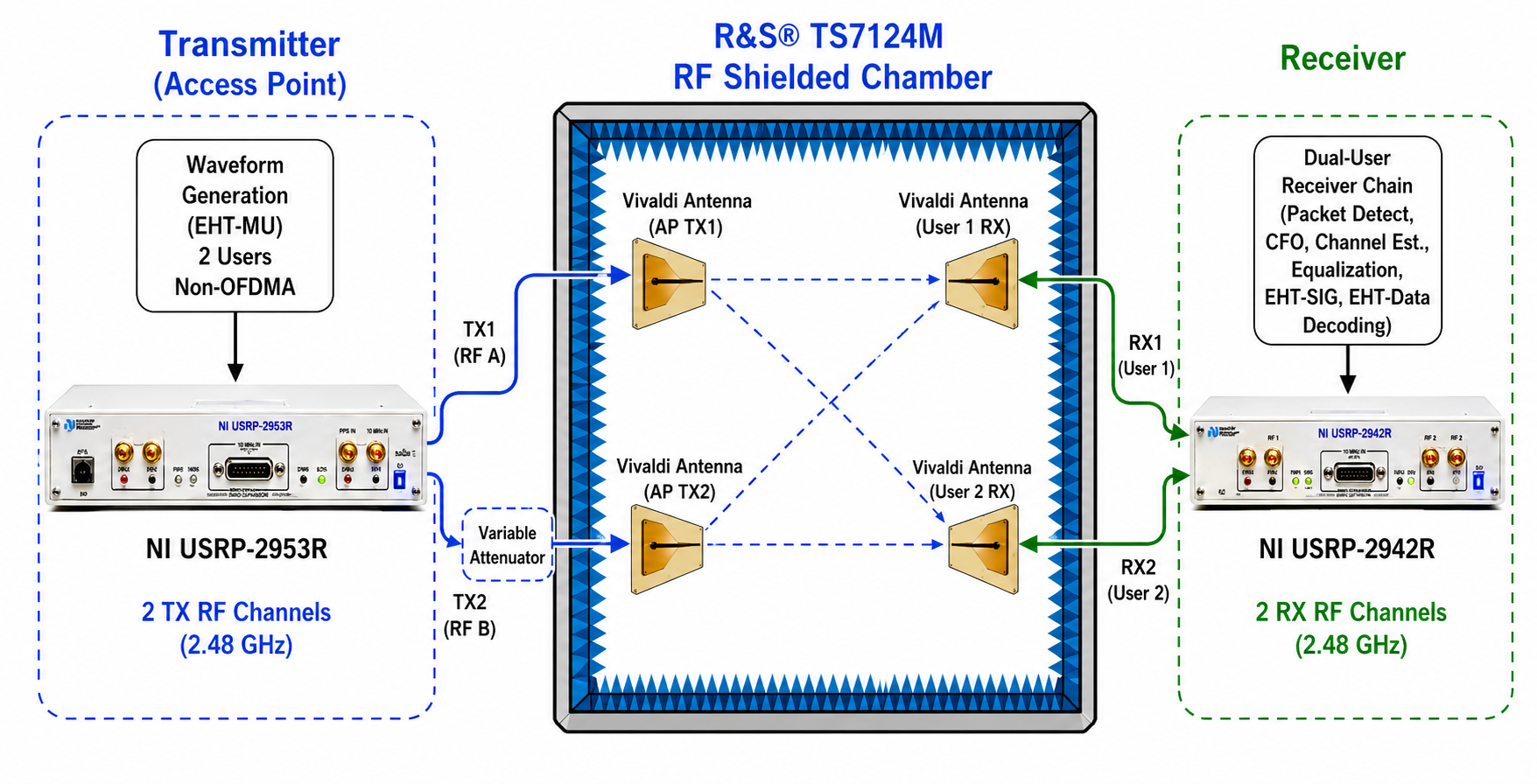}
    \caption{Controlled OTA measurement setup in a shielded enclosure for dual-user non-OFDMA IEEE 802.11be EHT-MU testing using two NI USRP-2953R and NI USRP-2942R radios.}
    \label{fig:setup}
\end{figure}

\begin{figure}[t]
    \centering
    \includegraphics[width=0.90\columnwidth]{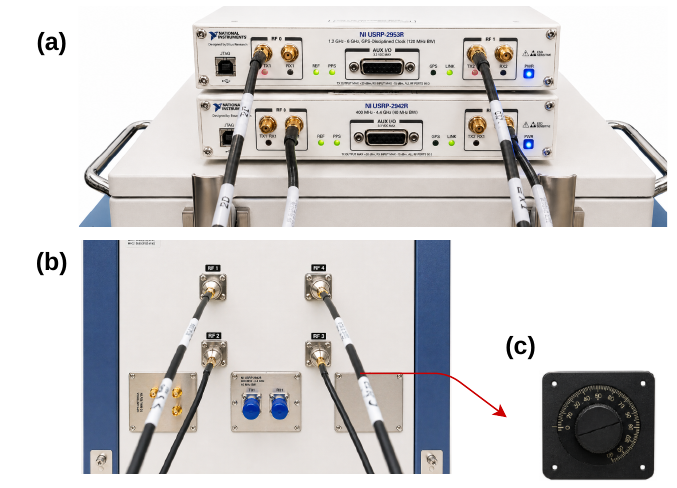}
    \caption{Experimental RF cabling setup: (a) NI USRP-2953R and NI USRP-2942R units, (b) RF interface panel of the shielded chamber, and (c) external variable attenuator.}
    \label{fig:rf_cabling_setup}
\end{figure}

Controlled OTA measurements were conducted using
NI USRP-2953R \cite{ni_usrp2953_specs} and NI USRP-2942R \cite{NI_USRP2942_specs} software-defined radios in an R\&S\textsuperscript{\textregistered}TS7124M RF shielded chamber \cite{rs_ts7124m_manual}. The NI USRP-2953R acts as the AP transmitter and generates the dual-user IEEE 802.11be EHT-MU waveform over two RF channels, while the NI USRP-2942R captures two synchronized receive branches and implements the dual-user recovery chain. The AP employs two transmit chains, while the two receive branches logically represent two single-antenna users with one spatial stream per user. Four identical R\&S TS-F24-V1 broadband Vivaldi antennas were installed inside the chamber and connected to the external USRP platforms through the rear RF feedthroughs (see Fig.\ref{fig:setup} and Fig. \ref{fig:rf_cabling_setup}). The printed tapered-slot antennas operate from \(0.7\)~GHz to \(14\)~GHz, provide linear polarization and \(50~\Omega\) input impedance. The use of four identical antennas ensured a fixed and repeatable broadband radiating geometry throughout the measurements while keeping the instrumentation outside the enclosure. A calibrated precision variable attenuator (Weinschel Engineering Model~974-C1-4-72, 6--120~dB) was inserted between the second AP RF output and the corresponding transmit antenna. The sweep creates a controlled imbalance between the two AP transmit chains. The shielded-chamber configuration suppresses external interference, reduces uncontrolled environmental variability, and improves repeatability across attenuation sweeps.


\subsection{Metrics}
The two main performance metrics are EVM and BER. The per-user BER is defined as
\begin{equation}
\mathrm{BER}_u = \frac{1}{L_u}\sum_{i=1}^{L_u}\mathbf{1}\{\hat{b}_{u,i}\neq b_{u,i}\},
\end{equation}
where \(L_u\) is the number of compared payload bits for user \(u\). EVM quantifies the deviation between the received and ideal constellation points \cite{mckinley2004evm,tek80211evm}. The RMS EVM is
\begin{equation}
\mathrm{EVM}_u = \sqrt{\frac{\sum_i |\hat{s}_{u,i} - s_{u,i}|^2}{\sum_i |s_{u,i}|^2}},
\end{equation}
where \(s_{u,i}\) and \(\hat{s}_{u,i}\) are the reference and equalized symbols for user \(u\). EVM is particularly useful for identifying quality degradation before hard-decision failure. At the packet stage, the success probabilities of U-SIG recovery, EHT-SIG common recovery, EHT-SIG user recovery, and overall packet recovery are also evaluated. The packet-recovery event is defined as the logical conjunction of packet detection, format detection, U-SIG recovery, EHT-SIG common recovery, and EHT-SIG user recovery. The waveform and receiver settings listed in Table~\ref{tab:waveform_params} were used throughout the measurement campaign.
\begin{table}[t]
\caption{Baseband waveform and measurement parameters.}
\label{tab:waveform_params}
\centering
\begin{tabular}{ll}
\hline
Parameter & Value \\
\hline
Channel bandwidth & CBW20 \\
Center frequency & 2.48 GHz \\
Number of users & 2 \\
Spatial streams per user & 1 \\
Total spatial streams & 2 \\
User MCS &  $[MCS2\;\;MCS2]$ \\
User~2 coding & LDPC / BCC selectable \\
Sweep variable & TX2 attenuation \\
MSDU length & 1500 bytes per user \\
MAC frame type & QoS Data \\
PSDU source & Valid MAC MPDUs with FCS \\
TX peak scaling & 0.35 \\
\hline
\end{tabular}
\end{table}

\subsection{Waveform Generation and Receiver Processing Chain}
A dual-user IEEE 802.11be EHT-MU baseband waveform was generated using the WLAN Toolbox with channel bandwidth set to CBW20 and two AP transmit antennas. Each user was assigned one spatial stream, so that the transmitted packet corresponds to a two-user downlink MU-MIMO transmission with one stream per user. The user payloads were created from valid MAC-layer QoS data frames. For each user, a random MAC service data unit (MSDU) of 1500 bytes was generated, encapsulated into a MAC protocol data unit, and converted into physical service data unit (PSDU) bits. These PSDU bits were then passed to the EHT waveform generator, ensuring that the transmitted packet contained standards-compliant MAC framing rather than arbitrary random payload bits. This enables subsequent MAC protocol data unit (MPDU) decoding and frame-check-sequence verification upon successful PHY recovery. The generated waveform was normalized and scaled to a controlled peak factor before being continuously transmitted from the AP USRP. We also log the transmit peak-to-average power ratio to verify that the waveform is not excessively clipped by the RF front end. At the receiver, the second USRP captures complex baseband samples from two receive channels. Packet recovery begins with front-end synchronization. Packet detection is performed using the legacy short training field, followed by coarse carrier-frequency-offset estimation. A refined packet timing offset is then obtained from symbol timing estimation, after which fine carrier-frequency-offset estimation is performed from the legacy long training field. The corrected waveform is subsequently rescaled using the received short-training-field power in an automatic gain control normalization step. Once synchronization is established, the receiver performs packet-format detection and identifies the captured waveform as EHT-MU. Channel estimation and noise-variance estimation are first obtained from the legacy long training field, which also provides an LLTF-based signal-to-noise-ratio estimate. The legacy signal field is then demodulated and equalized to recover the length information and verify successful legacy-header decoding. After this, the receiver processes the universal signal field and the EHT signal fields, including both the common and user-specific portions. Successful decoding of these fields determines whether the packet structure and user allocation information have been reliably recovered. After control-field recovery, the receiver processes the EHT-LTF and EHT-Data portions for each decoded user. Channel estimation is obtained from the EHT-LTF, followed by pilot-aided tracking, OFDM demodulation, noise estimation, frequency-domain equalization, and bit recovery. The recovered bits are then compared against the transmitted PSDU bits in order to compute the per-user BER. The recovered MPDU is also passed to the MAC decoder for frame validation and frame check sequence (FCS) inspection. In addition to BER, the receiver computes several signal-quality metrics for each user. These include EHT-Data EVM, noise variance, mean channel power, an estimated SNR based on channel-power-to-noise ratio, and an EVM-derived SNR. The equalized constellation, EVM per OFDM symbol, EVM per subcarrier, and estimated channel power spectrum are exported for post-processing. At the packet level, we also log synchronization status, packet format, coarse carrier frequency offset (CFO) and fine CFO, LLTF-estimated SNR, L-SIG EVM, U-SIG EVM, and the pass/fail status of U-SIG and EHT-SIG decoding. This provides a complete measurement chain from waveform synthesis to per-user payload recovery and packet-stage diagnostics.

\section{Results and Discussion}
The following results use the logged packet-stage and user-stage metrics to separate three effects: progressive degradation of the stream decoded as User~2, packet-level signaling robustness, and coding-dependent payload collapse.

\subsection{Per-User BER and EHT-Data EVM}
\begin{figure}[t]
    \centering
    \includegraphics[width=0.90\columnwidth]{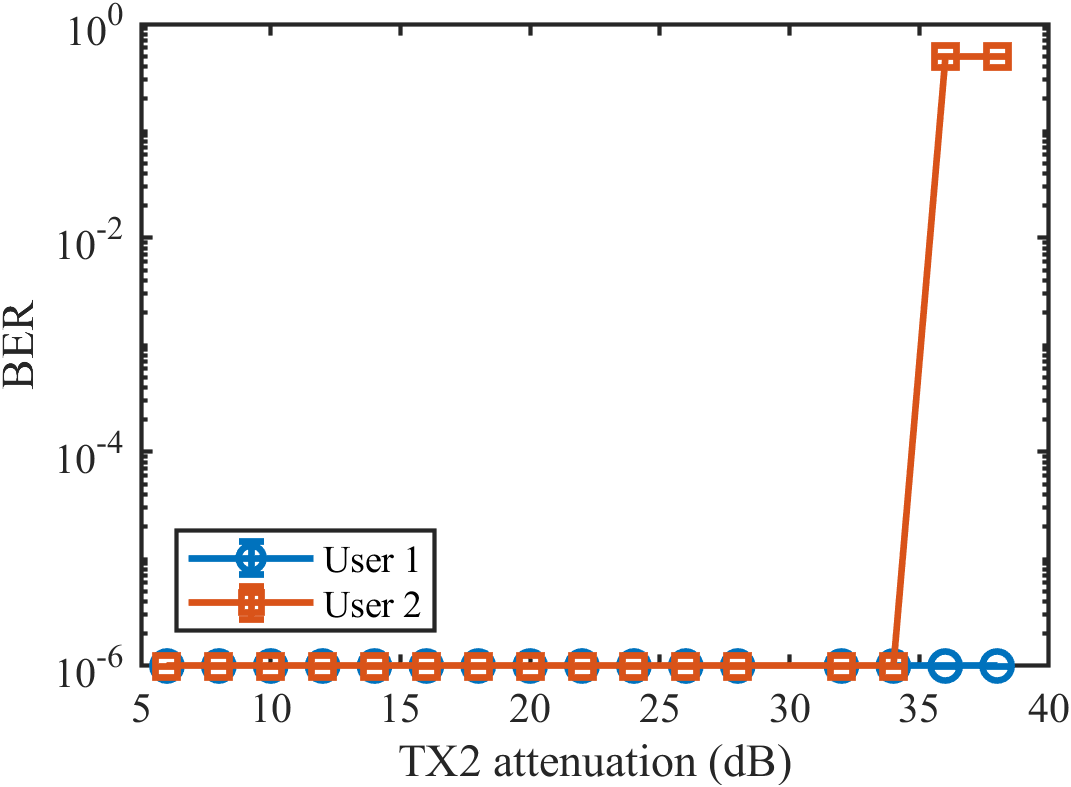}
    \caption{Per-user BER versus TX2 attenuation.}
    \label{fig:ber}
\end{figure}

\begin{figure}[t]
    \centering
    \includegraphics[width=0.90\columnwidth]{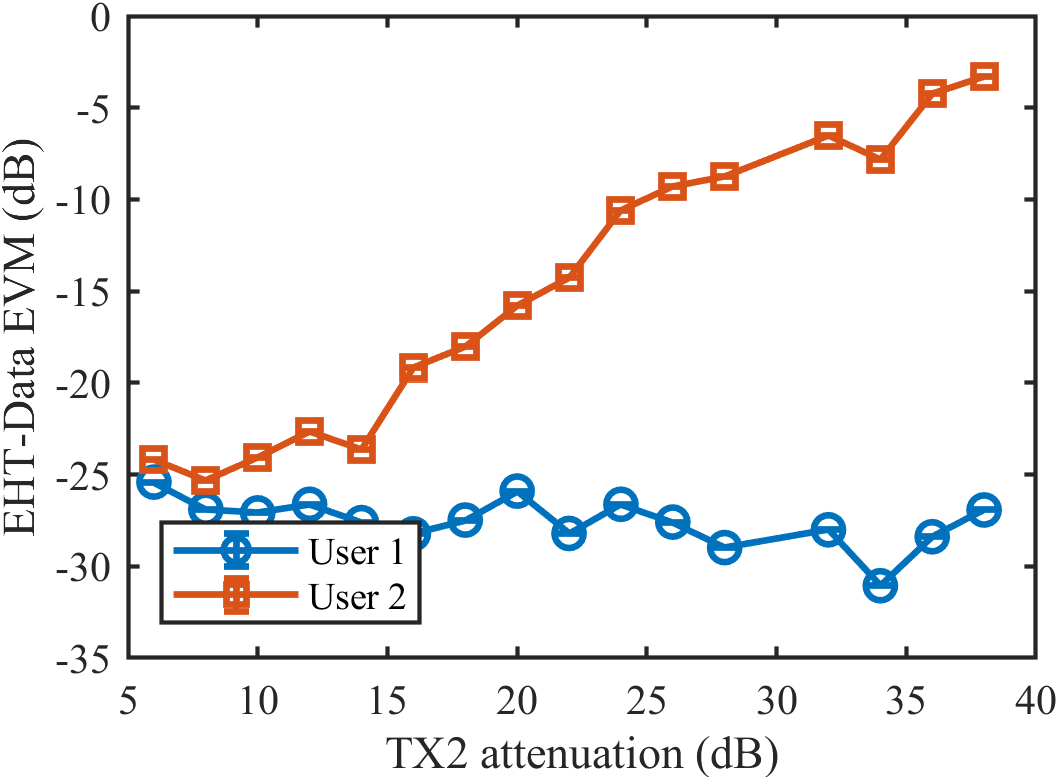}
    \caption{Per-user EHT-Data EVM versus TX2 attenuation.}
    \label{fig:evm}
\end{figure}

Fig.~\ref{fig:ber} shows the per-user BER as a function of TX2 attenuation. User~1 remains at the BER floor throughout the sweep, indicating that the stream decoded as User~1 remains robust under the tested transmit-chain imbalance. In contrast, the stream decoded as User~2 exhibits a clear threshold-like degradation behavior. For low and moderate TX2 attenuation values, User~2 BER remains at the floor, indicating reliable payload recovery. However, once the TX2 attenuation approaches the edge of the supported operating region, the BER increases abruptly and reaches a near-failure regime at the highest attenuation points. Fig.~\ref{fig:evm} shows the corresponding per-user EHT-Data EVM. In contrast to the threshold-like BER behavior, the EVM trend reveals a more gradual progression of signal-quality degradation. User~2 EVM worsens steadily as TX2 attenuation increases, whereas User~1 EVM remains comparatively stable over the same sweep. This indicates that the imposed transmit-chain imbalance maps primarily onto the stream decoded as User~2 in the measured geometry and receiver configuration. The EVM curve therefore, acts as an early indicator of deterioration, while the BER curve captures the final transition to unsuccessful payload recovery. The stability of User~1 further shows that the observed degradation is stream-dependent rather than a packet-global failure.
\subsection{Control-Field and Packet Recovery Success}
\begin{figure}[t]
    \centering
    \includegraphics[width=0.90\columnwidth]{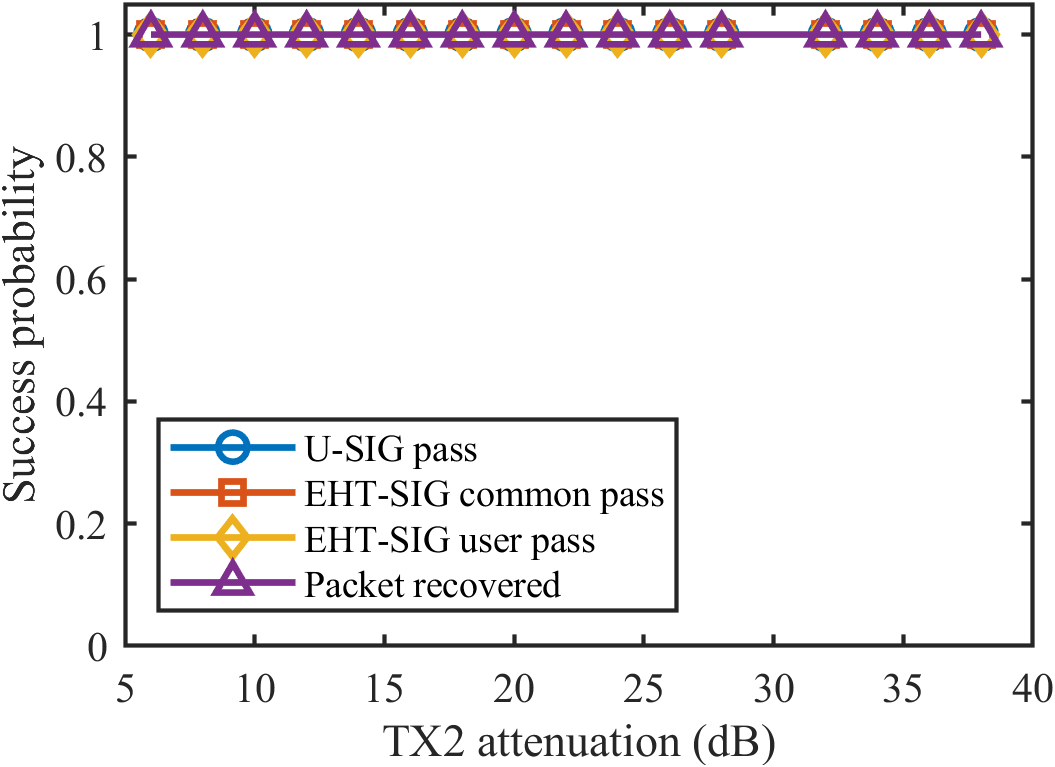}
    \caption{Control-field and packet recovery success probabilities versus TX2 attenuation.}
    \label{fig:control}
\end{figure}

Fig.~\ref{fig:control} plots the success probabilities of U-SIG recovery, EHT-SIG common recovery, EHT-SIG user recovery, and overall packet recovery as functions of TX2 attenuation. Over the tested attenuation range, all four probabilities remain close to one. This indicates that the shared control and signaling stages of the packet remain highly robust even as the User~2 payload stream is progressively degraded. Thus, the TX2 attenuation does not immediately cause packet detection failure, common-field decoding failure, or packet-stage signaling collapse. Instead, the dominant degradation appears after packet interpretation, in the EHT-Data recovery of the more sensitive stream. The close clustering of the four curves also suggests that the OTA link remains stable from the perspective of synchronization and header recovery. The controlled transmit-chain attenuation is not causing a global packet failure mechanism. The measurements indicate that the dominant impairment mechanism in the tested regime is stream-local payload degradation, not collapse of the common packet structure. This supports the interpretation that the measured link operates in a regime where user-specific data robustness is the limiting factor, rather than packet-level synchronization or signaling robustness.
\subsection{Coding Comparison: User~2 LDPC versus BCC}
\begin{figure}[t]
    \centering
    \includegraphics[width=0.90\columnwidth]{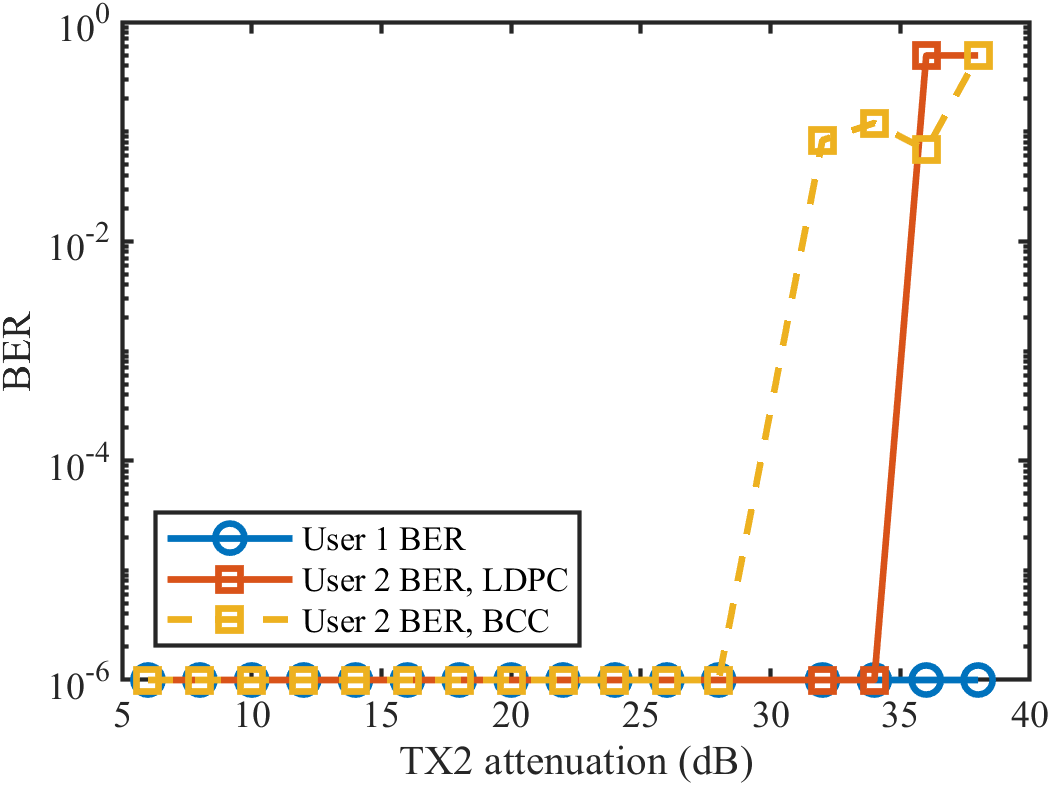}
    \caption{BER comparison for User~1, User~2 with LDPC, and User~2 with BCC under TX2 attenuation.}
    \label{fig:bercmp}
\end{figure}

\begin{figure}[t]
    \centering
    \includegraphics[width=0.90\columnwidth]{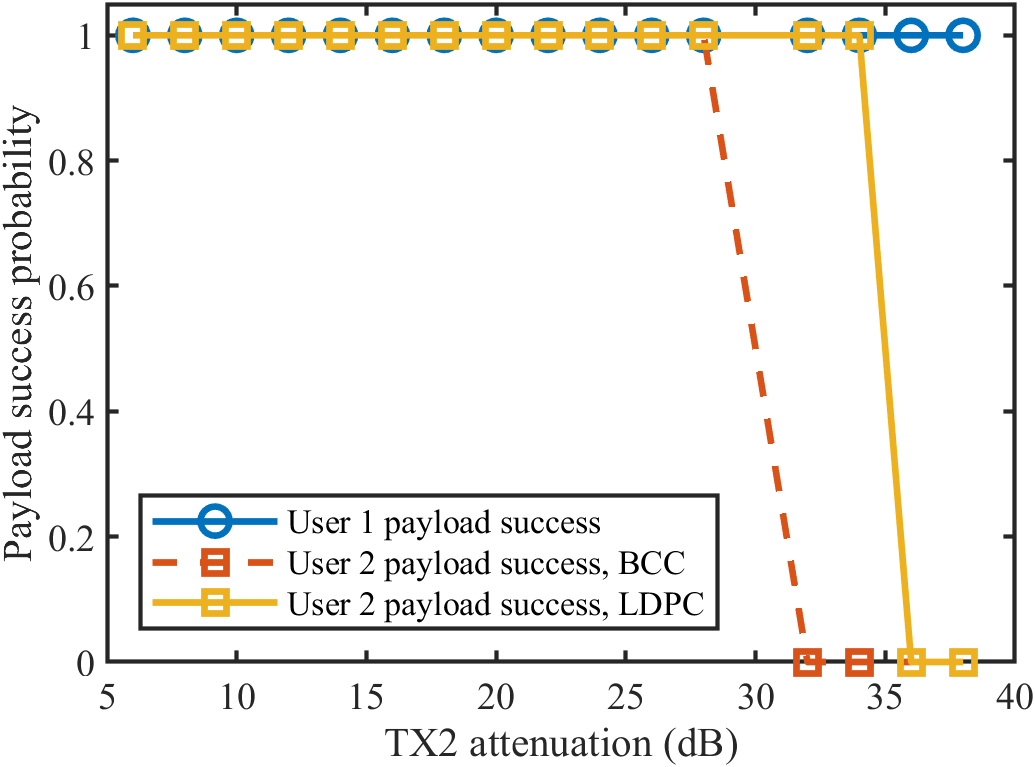}
    \caption{Payload-success comparison for User~1, User~2 with LDPC, and User~2 with BCC under TX2 attenuation.}
    \label{fig:payloadcmp}
\end{figure}

To quantify coding sensitivity for the more sensitive recovered stream, Fig.~\ref{fig:bercmp} compares BER for three curves: User~1 BER, User~2 BER with LDPC, and User~2 BER with BCC. User~1 remains at the BER floor less than \(10^{-6}\) across the full measured TX2 attenuation range. In contrast, User~2 with BCC begins to degrade at approximately \(30\)~dB TX2 attenuation, while User~2 with LDPC remains at the BER floor until approximately \(35\)~dB TX2 attenuation. This indicates an LDPC advantage of roughly \(5\)~dB in TX2-attenuation margin for the stream decoded as User~2.
Fig.~\ref{fig:payloadcmp} shows the corresponding payload-success comparison. User~1 maintains payload success essentially equal to one across the full measured attenuation range. In contrast, User~2 exhibits a clear coding-dependent transition. With BCC, payload success begins to collapse at approximately \(30\)~dB TX2 attenuation, whereas with LDPC, successful payload recovery is maintained until approximately \(35\)~dB. This result is important because payload success provides a direct end-to-end indicator of whether the recovered User~2 payload remains usable, complementing the gradual EVM degradation and threshold-like BER transition. The figure shows that the coding choice changes the attenuation point at which reliable payload delivery is lost. In the measured setup, LDPC therefore extends the usable operating region of the more sensitive stream by approximately \(5\)~dB relative to BCC. The payload-success curves also reinforce the interpretation that the dominant impairment mechanism remains stream-local. Even when User~2 payload success collapses, User~1 continues to exhibit essentially error-free operation, consistent with the earlier BER and control-field results.

\subsection{Coding Comparison via User~2 Subcarrier-Level EVM}
\begin{figure}[t]
    \centering
    \includegraphics[width=0.90\columnwidth]{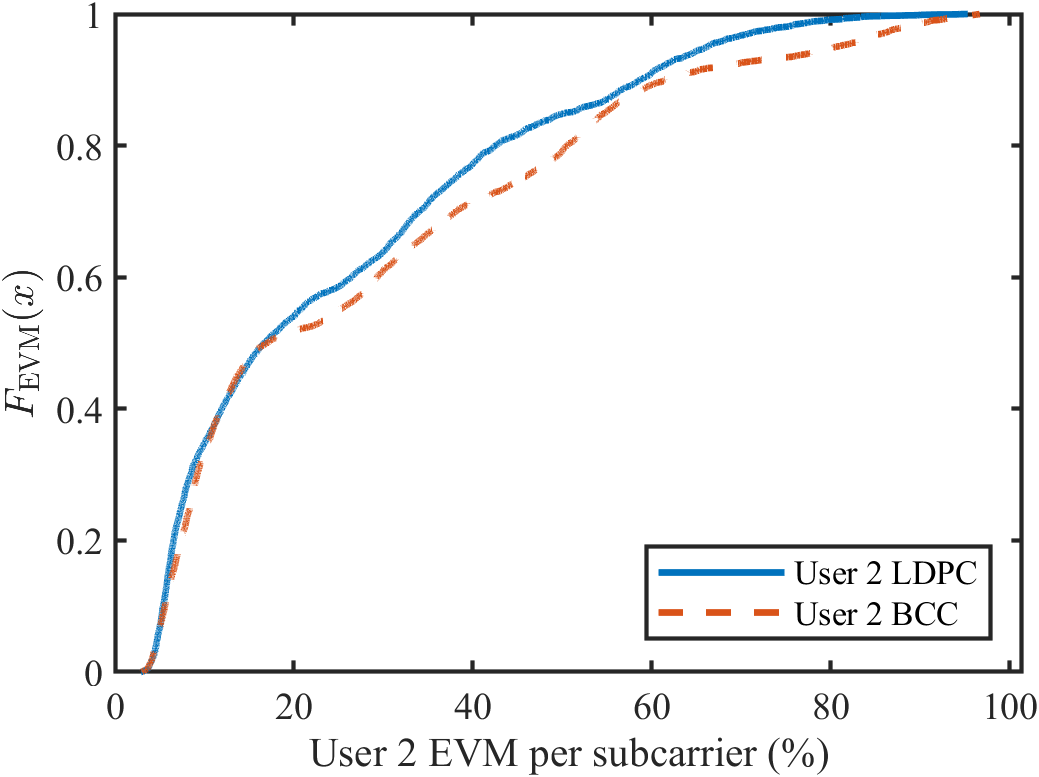}
    \caption{Empirical cumulative distribution functions of User~2 EVM per subcarrier for low-density parity-check (LDPC) and binary convolutional coding (BCC). The BCC curve is shifted to the right, indicating a larger fraction of high-EVM subcarrier observations.}
    \label{fig:cdf_evm_subcarrier_user2}
\end{figure}

\begin{figure}[t]
    \centering
    \includegraphics[width=0.90\columnwidth]{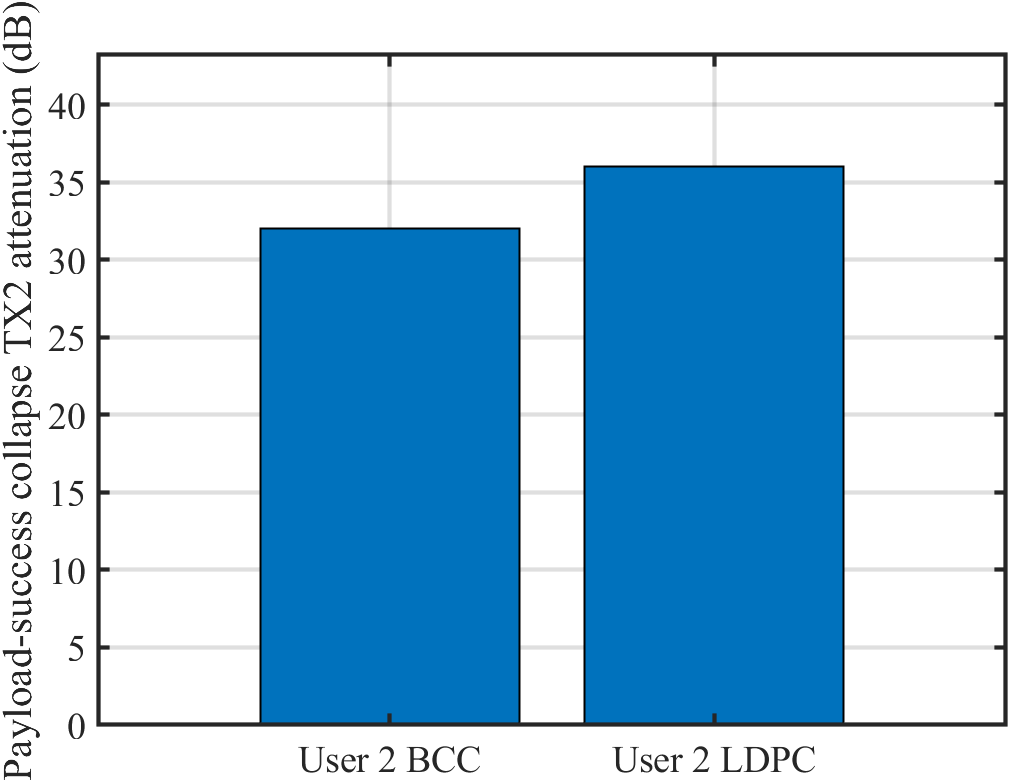}
    \caption{Payload-success collapse attenuation for User~2 with BCC and LDPC under TX2 attenuation. LDPC shifts the collapse point to a higher TX2 attenuation.}
    \label{fig:payload_collapse_margin}
\end{figure}

To further examine the coding-dependent robustness of the more sensitive stream, Fig.~\ref{fig:cdf_evm_subcarrier_user2} shows the empirical cumulative distribution functions (CDFs) of User~2 subcarrier-level EVM for the LDPC and BCC cases. The BCC case has a larger proportion of high-EVM subcarrier observations. This trend is also reflected in the quantile statistics, for which the BCC-minus-LDPC gap is positive at all quantiles and becomes particularly pronounced in the upper-middle portion of the distribution. This suggests that the earlier BER and payload-success breakdown observed for BCC is associated with a heavier tail of degraded subcarrier observations.
The combined results support a consistent interpretation. First, TX2 attenuation produces a smooth degradation in the EHT-Data EVM of the stream decoded as User~2. Second, User~2 BER rises significantly only at higher TX2 attenuation values. Third, User~1 remains stable across the sweep. Fourth, common signaling fields remain recoverable over the tested operating range. Finally, replacing User~2 BCC with LDPC delays BER and payload-success collapse by approximately \(5\)~dB of TX2 attenuation. These observations show that the measured regime is dominated by stream-dependent payload degradation rather than packet-global failure. Fig.~\ref{fig:payload_collapse_margin} shows that User~2 with LDPC maintains payload recovery to a higher TX2 attenuation than User~2 with BCC, providing a compact visualization of the coding-margin improvement. 
These results are useful for both receiver evaluation and scheduler-oriented studies. In practical terms, they show that a transmit-chain imbalance does not necessarily produce packet-global failure. Instead, the impairment appears primarily as degradation of one recovered user stream while the other stream and common signaling remain robust. 

\section{Conclusion}
This paper presented a controlled OTA characterization of dual-user non-OFDMA IEEE 802.11be EHT-MU transmission in a shielded enclosure. The measurements show that increasing transmit-chain imbalance primarily degrades the stream decoded as User~2, evidenced by worsening EHT-Data EVM and rising BER, while the stream decoded as User~1 and the common packet signaling remain robust over the tested range. Thus, the imposed transmit-chain imbalance produces a stream-dependent payload degradation in the measured configuration rather than packet-global signaling collapse. Quantitatively, User~1 remained at the BER floor less than \(10^{-6}\) across the full measured TX2 attenuation range, while User~2 with BCC exhibited BER and payload-success degradation at approximately \(30\)~dB TX2 attenuation. In comparison, User~2 with LDPC maintained BER-floor performance and payload success until approximately \(35\)~dB, corresponding to an attenuation-margin gain of about \(5\)~dB. The empirical CDF analysis further shows that the BCC case is associated with a heavier high-EVM tail, providing a distribution-level explanation for the earlier payload-success collapse. Future perspective of the work includes extension of the methodology to wider bandwidths, additional MCS values, different stream mappings, and OFDMA configurations.

\balance
\bibliographystyle{IEEEtran}
\bibliography{references.bib}

\vfill

\end{document}